# Generative AI and News Consumption: Design Fictions and Critical Analysis

JOEL KISKOLA

Faculty of Information Technology and Communication, Tampere University, Tampere, Finland. joel.kiskola@tuni.fi

HENRIK RYDENFELT

Faculty of Social Sciences, University of Helsinki, Helsinki, Finland. henrik.rydenfelt@helsinki.fi

THOMAS OLSSON

Faculty of Information Technology and Communication, Tampere University, Tampere, Finland. thomas.olsson@tuni.fi

LAURI HAAPANEN

Faculty of Humanities and Social Sciences, University of Jyväskylä, Jyväskylä, Finland. lauri.m.haapanen@jyu.fi

NOORA VÄNTTINEN

Department of Industrial Engineering and Management, Aalto University, Espoo, Finland. noora.vanttinen@aalto.fi

MATTI NELIMARKKA

Centre for Social Data Science, CSDS, University of Helsinki, Helsinki, Finland. matti.nelimarkka@helsinki.fi

MINNA VIGREN

Faculty of Social Sciences, LUT School of Engineering Sciences, Lappeenranta, Finland. minna.vigren@lut.fi

SALLA-MAARIA LAAKSONEN

Faculty of Social Sciences, University of Helsinki, Helsinki, Finland. salla.laaksonen@helsinki.fi

TUUKKA LEHTINIEMI

Centre for Consumer Society Research, University of Helsinki, Helsinki, Finland. tuukka.lehtiniemi@helsinki.fi

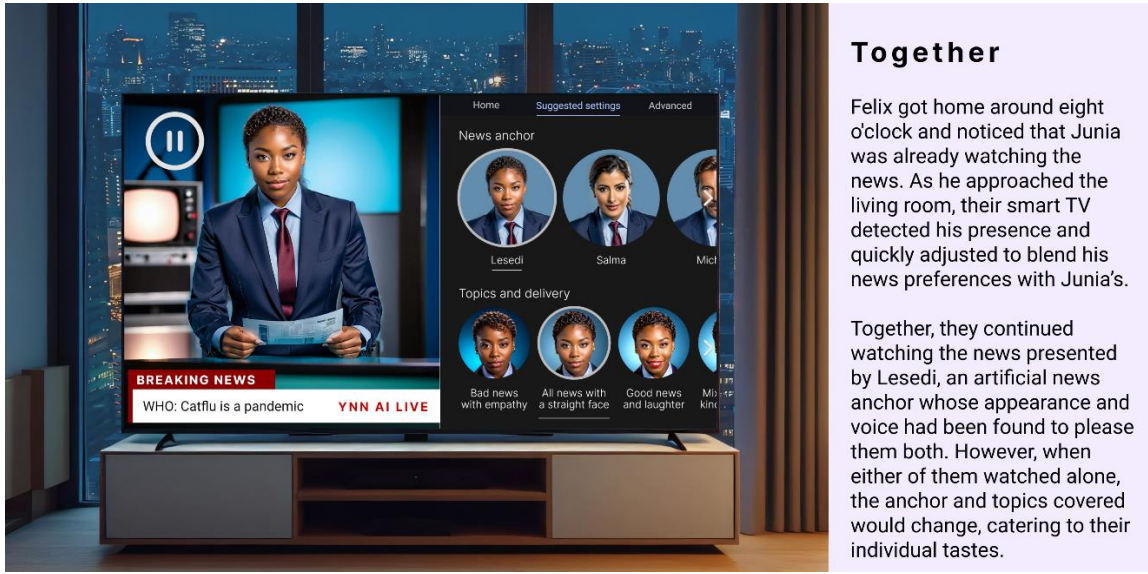

Figure 1: Design fiction 5 (out of 6): *Together*. A news broadcast delivered by a virtual news anchor, customized to appeal to multiple viewers' preferences. The fiction embodies workshop ideas of personalized news presenters and an AI that selects relevant news for the user.

The emergence of Generative AI features in news applications may radically change news consumption and challenge journalistic practices. To explore the future potentials and risks of this understudied area, we created six design fictions depicting scenarios such as virtual companions delivering news summaries to the user, AI providing context to news topics, and content being transformed into other formats on demand. The fictions, discussed with a multi-disciplinary group of experts, enabled a critical examination of the diverse ethical, societal, and journalistic implications of AI shaping this everyday activity. The discussions raised several concerns, suggesting that such consumer-oriented AI applications can clash with journalistic values and processes. These include fears that neither consumers nor AI could successfully balance engagement, objectivity, and truth, leading to growing detachment from shared understanding. We offer critical insights into the potential long-term effects to guide design efforts in this emerging application area of GenAI.

## 1 INTRODUCTION

The field of Human-Computer Interaction (HCI) is increasingly engaged in critical discussions on the integration of Artificial Intelligence (AI) into various application areas, reflecting the rapid advancements in AI technologies and their complex ethical, social, and technical challenges (e.g., [12, 26, 48–50, 68, 85]). HCI researchers are particularly concerned with the implications of AI on user autonomy, privacy, and broader societal impacts, calling for more responsible and human-centered AI development [64, 79]. However, while significant attention has been devoted to AI's role in various domains, the media industry—a sector with profound societal influence—has received comparatively less focus, especially regarding how AI might be readily utilized in daily interactions around news consumption.

Over the past decade, news consumption has been drastically shaped by digital platforms, social media, and tailor-made news applications [31, 62]. In this context, prior HCI research has explored the design of news applications (e.g., [14, 18, 60, 90, 103]), but AI as a new design material has received little attention in academic research. At the same time, also media and journalism research has paid little attention to AI in news consumption, focusing on AI in the production of news instead [20, 41, 58, 66, 84, 94, 100]. This oversight is notable as AI has nonetheless been embedded in news applications in various forms, ranging from AI-driven summarizers to recommender systems that personalize news feeds, and news applications are marketed as AI-driven /enhanced/powered. For example, the developers of applications like Ground News [33], Fuse [27], and Listen2.AI [54] aim to revolutionize news consumption through AI-driven personalization and political balancing of news content.

To address these gaps, our study adopts a critical approach to examine the potential and threats of AI-powered news applications. We utilized ideation workshops and design fiction (see e.g., Figure 1) to create scenarios exploring plausible AI-enhanced news consumption futures. Design fiction was chosen as the key method as it allows for the exploration and contrasting of several alternative futures, where both the development of the fictions and the provoked discussions allow complementary forms of inquiry [5, 91, 92]. By imagining and crafting speculative futures, design fiction opens up space for reflecting on the societal implications of emerging technologies (ibid.). We then conducted workshops with experts in journalism, media, and technology to analyze the fictions so as to address the research question: *What are the possible societal implications of adopting AI-enhanced news applications?* We approach this question by examining the assumptions made within the fictions of technology, society, and people, as well as the potential societal consequences of the adoption of the imagined technologies.

We make two contributions to HCI research: (1) we present speculative design fictions that explore the future of AI-driven news consumption and thus help outline both desirable and undesirable directions for this nascent application area of AI, and (2) we offer critical insights and concerns regarding the adoption of AI-enhanced news applications and services. Based on the findings, we underscore the importance of integrating interdisciplinary expertise into the development of AI news applications and being mindful of various recognized risks. Thus, this work enhances the knowledge base of HCI researchers and promotes ethical considerations in the development of future AI news technologies.

## 2 BACKGROUND AND RELATED WORK

### 2.1 Emerging Technologies in News Production and Consumption

The media industry appears to be keenly aware of the transformative power that AI technologies could wield for newsroom activities [34, 88]. In research on AI in media and journalism, the focus has been on the production and dissemination of content and related issues pertaining to editorial control and journalistic values [20, 24, 41, 58, 81, 84, 94, 100], audience accounts on journalists using AI [23], and informed the design of AI systems for journalists [45, 66]. Moreover, researchers have paid close attention to the ways in which journalistic media has had to conform to the operation of the algorithmic technologies of social media platforms to reach their audiences [16, 88]. However, while recent studies have briefly explored audiences' experiences and views of the use of algorithms and AI in the production and distribution of content by news media (e.g., [38, 94]), research has paid little attention to the role of AI from the audience's perspective, and we are not aware of any studies exploring AI-enhanced news applications for consumers.

In this research, the broad notion of AI is mainly used to refer to large language models (LLMs) or other forms of Generative AI (GenAI). GenAI, an advanced AI branch, produces text, images, and other data forms [61]. These systems can present information visually through graphs and generate images and videos to enhance text. Popular GenAIs like ChatGPT demonstrate impressive natural language generation, mimicking human writing and conversation [29]. As LLMs continue to evolve, they can potentially increasingly handle tasks traditionally associated with journalism. LLMs already represent significant progress over earlier systems, which relied on structured data for news generation [19]. A recent study suggests that GPT-4 surpasses its predecessor 3.5 in fact-checking news, albeit still falling short compared to human fact-checkers [9]. All in all, from a technical perspective, it appears that GenAI features such new computational capabilities that would fit this application area, which is characterized by language processing and immediacy. These observations serve as the key premise and motivator of the present exploratory work.

Further, our work conceptually relates to previous explorative research in the field of HCI on consumer news applications not featuring GenAI. In addition to earlier user experience studies on news apps (e.g., [14, 103]), for example, Sotirakou & Mourlas [90] explored gamifying mobile news applications. They presented a design of a mobile reader application, implementing game elements, like rewards for reading, so that the user experience would be improved and users would read more news. This relates conceptually to some of our design fictions, where it is proposed that users could do new, somewhat playful things, such as using AI to discover more about the news or customize the news. Munson et al. [60] studied a browser extension that provided users with feedback on the political leanings of their reading habits, encouraging more balanced news consumption. They found that showing feedback on the political balance of the news consumption led to a modest move toward politically balanced exposure. This relates conceptually to one of our design fictions, where it is proposed to give users full control over the political aspects of news they consume.

## 2.2 Design Fiction as a Research Method

Design fiction (DF), initially introduced by Bruce Sterling in his 2005 book "Shaping Things" [91] and further refined by Julian Bleecker in 2009 [4], employs speculative design to critically examine and reflect on the future by using world-building and narrative strategies. This approach involves creating fictional prototypes within imagined worlds to provoke thought about future possibilities, the desirability, and consequences of certain technologies, and to place viewers in different conceptual spaces temporarily [35, 102]. DF envisions future scenarios by creating "fantasy prototypes" that spark discussions on concept designs without needing to be built, addressing contemporary issues and the impacts of science and technology [6, 17].

DFs can take various forms, including narratives, films, physical prototypes, text, images, audio fragments, video, objects, or experiential prototypes, depicting both utopian and dystopian futures [6, 56]. These are either co-created with participants or participants are asked to give feedback on pre-designed fictions. As an evolving method, DF is flexible and open to creative and innovative applications [52, 56, 77, 92]. Moreover, DF is used in HCI research and computer science education [44, 57, 75] as there is an increasing need to investigate the potential future harms and consequences of current technologies [44, 70].

In the present study, we create DFs consisting of images illustrating possible AI-enhanced news applications and short stories about their use. We were inspired to create these types of artifacts by previous DFs, such as *BreakApp* by Blythe & Encinas [7] and *Jobnettrace* by Moller et al. [39]. For example, *BreakApp* is a DF around a mobile divorce app, consisting of a figure illustrating the imagined divorce app, a short story (epilogue) about its use, an advert, "Found a new you?", and a magic realist story "One hundred years of solitude." We follow a similar combination of a figure and story. *Jobnettrace* is a realistic-looking DF of a platform and app where unemployed individuals make themselves eligible for support through sharing data, presented in the form of a short video. While the present study does not utilize video, in this study DFs are likewise intended to appear relatively realistic and normal. Further, we will note some coincidental resemblances to other DFs and science fiction works when presenting our DFs.

Building on the tradition of employing DFs to provoke critical thinking and reflection, previous HCI research has utilized DFs to engage participants and gather insights [1, 43, 47, 75, 96, 101]. To give some examples, Ahmadpour et al. [1] organized two focus group sessions with older adult participants to gather feedback on DF of future well-being technologies. Rezwana & Maher [75] utilized DF in a survey and focus group with experts to explore user perspectives around ethical issues in human-AI co-creation. Wong et al. [101] showed technology professionals collections of DFs on surveillance technologies, which elicited reflections and discussions on values.

## 3 METHODOLOGY

The research process included two main phases (Table 1) to, first, create thought-provoking DFs and, second, gather expert insights to address the main research question. Each phase comprised several activities and concrete tasks to generate and analyze elements of the DFs. The process followed a structured approach to ideating, developing and evaluating DFs based on expert understanding.

The fictions contribute to the RQ with experts' beliefs on what technologies could plausibly exist from the perspectives of technical feasibility, consumer needs, and trends in digital service design. However, they are not intended to offer normative guidance on future applications [6, 17]. The fictions diverge and accelerate from the present to the point of discomfort, so that they would trigger reflection among viewers [73]. Accordingly, they are also used as stimuli in phase 2 to get experts to raise issues and make noteworthy observations. Additionally, the fictions serve to connect the experts' comments to tangible artifacts, enabling designers to more easily grasp expert insights. Principles of research ethics, including informed consent, were followed in all stages.

Table 1: Research design and methodology

|  | Activity | Tasks |
|---|---|---|
| Phase 1. (Crafting design Fictions) | Ideation Workshops | Generating large numbers of alternative ideas with experts based on imagined future capabilities of AI systems and what might be desirable for consumers |
|  | Review and Expansion of the Ideas | Transcribing and reviewing recordings. Expanding workshop ideas and identifying recurring, core ideas |
|  | Creation of Design Fictions | Creating DFs based on the generated ideas, particularly the recurring core ideas |
|  | Selection and Adjustment of Design Fictions | Narrowing down the number of DFs from 12 to 6, adjusting them to better highlight the recurring core ideas |
| Phase 2. (Expert analysis) | Survey and Workshops | Presenting the fictions to experts and gathering their feedback on AI applications' impacts, assumptions, etc. |
|  | Co-Writing and Analysis | Transcribing and reviewing recordings, organizing insights, and collaborating with experts to expand on key thoughts about AI and DFs |

## 3.1 Phase 1: Ideation and Development of Design Fictions

*3.1.1 Participants*

The ideation workshops involved the design team (authors 1–5) and ten other researchers. The other activities involved only the design team. The design team's areas of expertise were human-computer interaction, design, journalism, media and communication, and philosophy. The ten other researchers' areas of expertise spanned journalism, media and communication, human-computer interaction, and political science, with a shared interest in technology's impact on society, digital culture, and consumer behavior.

*3.1.2 Ideation Workshops*

In the first phase, two ideation workshops were held to brainstorm innovative ideas for AI in media consumption. The workshops, each lasting two hours, were facilitated by the first author and involved the participation of 6 to 7 researchers, including 3 members of the design team. Sessions began with reflecting on desires and expectations for AI's role in media, followed by collaborative brainstorming in small groups. The groups generated ideas for AI-enhanced news applications, both positive and negative, and briefly described them on paper. Each group included researchers with different areas of expertise to support their ideation and consideration of different perspectives.

To spark creativity, the participants were provided four depictions of different news consumers, PLEX cards [55], and the Tarot Cards of Tech [2] as inspiration material, as well as a brief list of present-day AI capabilities at a high level of abstraction. The depictions of various news consumers showcased diversity across age, gender, ethnicity, native language, visual ability, and news consumption habits and interests. These depictions were fictional and crafted by the design team. The list of AI capabilities included: generating and transforming text, audio, image, and video content; browsing websites and retrieving information; drawing insights from audio, image, and video content; recognizing emotions; recommending content; and interpreting contextual and ambiguous cues. The materials were chosen to address key design aspects: user empathy (news consumer depictions), playful engagement (PLEX cards), ethical foresight (Tarot Cards of Tech), and technical feasibility (AI capabilities). Together, these probes offered a well-rounded foundation for generating both optimistic and cautionary AI-driven news solutions.

*3.1.3 Review and Expansion of Ideas*

After the workshops, the first author analyzed the produced ideas and audio recordings to derive relevant elements for the fictions. This resulted in a collection of 20 unique design concepts, which laid the foundation for further fiction development. Recognizing that certain concepts were underrepresented, the first author spent a week expanding on the existing ideas, documenting 23 additional concepts. They did this by creatively applying different twists to the presented concepts. For example, while the participants mentioned content personalization, they did not mention personalizing video content for multiple simultaneous viewers, perhaps because of the limited time they had in the workshops.

Next, the first and second authors jointly categorized the concepts roughly based on the traditional stages of news production: gathering, assessing, creating, and presenting [66]. For example, an AI functionality of gathering a set of news articles for the user, but not modifying them, was considered to relate to the "gathering" stage. It was also decided that an additional "interactivity" should be introduced because the ideas included AI enhancing media interactivity, acting as a "discussion partner," fostering audience engagement with journalists, enabling sharing of AI-generated content, and customizing content to suit group preferences and understanding levels. This categorization was done to better understand the recurring, core ideas, and it further laid the groundwork for developing different DFs.

*3.1.4 Creation of Design Fictions*

Following the review and expansion of ideas, authors 1–4 agreed that the first author would develop DFs based on the documented ideas without the involvement of the rest of the research team to provoke thoughtful discussions among the rest of the research team to enable fresh viewpoints from the rest of them.

The 12 DFs were created through iterative reflection and engagement with empirical data as a foundation for design, together with a practice-based approach where the design was not solely about crafting objects but about exploring possibilities through advanced tools [30, 46, 105]. First, the previously created categorization of the documented ideas was used to help select different ideas, with those mentioned frequently by the workshop participants being prioritized and potential combinations for a single DF being considered. For example, the idea of using AI to transform content from one format into another (e.g., text to video) was combined with the ideas of using AI to summarize and expand on content and using AI to simulate a famous person commenting on the news (see DF6 under Results). Additionally, the news consumption practices of the ideation workshop participants were considered. Next, a concise yet vivid narrative was drafted, inspired by the chosen ideas and the stories told by the participants. The app designs were then refined using tools like Figma. Generative AI technologies were leveraged to both enhance the creative process and give the DFs presenting AI-generated content a plausible look and feel (e.g., DF2; Figure 3). ChatGPT was used to refine story ideas, ensuring consistency and coherence in the narratives. DALL-E and Stable Diffusion XL were utilized to generate custom visuals (depictions of users, news article images, and comic strip panels). With each DF, consideration was given to how it could reflect the variety of elements present in the documented ideas, with experimentation occurring with elements like speculativeness, dystopian or utopian themes, imagined users, motivations, contexts of use, and whether the focus was on user interface, app, or service. This process resulted in DFs that featured most of the documented concepts, particularly those that were mentioned recurringly.

*3.1.5 Selection and Adjustment of Design Fictions*

After completing the 12 fictions, the first author sought feedback from the other members of the design team by sending them a document with the fictions and probing questions. This was to prompt their reflection, benefit from their expertise, and explore the potential for future expert feedback through a survey.

After the members of the design team had individually thought about the fictions, the design team convened to discuss the next steps. The team decided to focus on near-future apps and services appealing to audiences, centering on content generation and being somewhat provocative. The team also selected concepts and features to highlight in the fictions, selecting mostly concepts that workshop ideas recurrently featured, for example, personalization and content transformation (Table 3 under Results). Additionally, the team decided to modify some of the fictions and combine others to better highlight the selected concepts and features. The six DFs presented in this paper resulted from this process. Specifically, concepts frequently highlighted during ideation workshops, such as personalization and content transformation, were prioritized (see Table 3 in the Results section). Key societal needs and speculative uses of AI in news were also emphasized: e.g., mental health considerations (DF1), accessibility for children (DF2), contextualized news (DF3, DF1), community-driven editorial review (DF3), user control over political, religious, and other content aspects (DF4), support for social consumption behaviors (DF5), and tools for news commentary and analysis (DF6, DF1).

## 3.2 Phase 2: Expert Analysis and Feedback

### 3.2.1 Survey and Workshops

In the second phase, the first author facilitated two semi-structured DF analysis workshops involving the 4 other members of the design team and 4 additional experts. 2 of them are women, 5 are men, and 1 is non-binary. All are Finnish. This phase included an open survey, analysis workshops, and a co-writing stage. Most of the workshop participants (henceforth "experts") had a background in different fields of social sciences. Also, most of them had been conducting studies on AI and digital technologies from a critical perspective. A few also had a background in journalism and communication. The design team were acquainted with the experts, and most of the experts knew each other. Table 2 details the experts' backgrounds. All of them later became co-authors of this study.

Table 2: Design fiction analysis workshop expert participant information

| Expert ID | Expertise |
| --- | --- |
| #1 (L.) | Journalist for 10 years. Journalism & Media Studies for 13 years. |
| #2 (S.) | Communication Studies for 16 years. Journalism & Media Studies, Consumer research, Political Science, and Social Computing for some years. |
| #3 (M.) | Media Studies for 16 years. Science and Technology Studies for 8 years. Some years of experience in Human-Computer Interaction, Journalism, and Media Education. |
| #4 (H.) | Philosophy for 20 years. Journalism, Media and Communication Studies for 10 years. Some experience in Journalism and Media Studies. |
| #5 (T.) | Computer Science, Human-Computer Interaction, and Design for 17 years. Some experience in IT Ethics and Science and Technology Studies. |
| #6 (M.) | Political Science for 15 years. Computer Science, Media & Communications for some years. |
| #7 (T.) | Economic Sociology and Economics for 16 years. Some years of experience in Electrical Engineering. |
| #8 (N.) | Some experience in Organizational Studies. Background in Control, Robotics and Autonomous Systems. |

In the survey stage, experts outside the design team (see #2, #3, #6, and #7 in Table 2) were provided with a presentation of the six DFs and asked to spend 2-3 hours answering four questions about them within two weeks, focusing on their initial reactions, audience needs, and potential impacts on journalism and society. After this, the first author facilitated two four-hour DF analysis workshops with 4–5 participants in each, some also being members of the design team. The workshops involved detailed discussions on three DFs per session. At the start of the workshops, participants shared their first impressions of the fictions and identified potential underlying issues. The facilitator then guided the discussion with questions such as: "Why might the solutions depicted in the fiction be inappropriate or inadequate?" "What should be changed in the application to make its use acceptable and meaningful?" "Considering various development paths and trends, how should the world change for fiction to be realistic?" and "If we think of journalism, its audience, technology developers, and the surrounding society as a socio-technical whole, what kind of impacts could the given technology have on different parts of the system?"

### 3.2.2 Data Analysis and Co-Writing

The data were analyzed and the paper presenting the research results was co-authored in four steps. First, the workshop recordings were transcribed by the first author, with assistance from Whisper, a machine-learning model. Second, both the pre-workshop survey data and transcriptions were analyzed by grouping together similar responses and issues raised. Third, the initial analysis was discussed, and it was decided that the analysis in this paper should focus on critical thoughts related to the assumptions made by the design team in crafting the DFs and the risks of the presented AI applications. Fourth, data were extracted by the first author to meet this aim and were expanded upon in collaboration with the experts who had participated in the workshops. This resulted in the following findings.

## 4 RESULTS

### 4.1 Design Fictions

The following presents the DFs developed in our study. The DFs do not aim to determine how certain things will inevitably come to be or to suggest how things should be but rather to stimulate discussion and reflection on future possibilities and alternatives [4, 91, 92]. They reflect the beliefs of ideation workshop participants and the diverse group of experts (Table 2) on what kind of AI-enhanced news applications could exist from the point of view of speculated future capabilities of AI and consumer desires. Table 3 outlines a categorization of high-level features recurring in the DFs, followed by detailed descriptions of each DF.

Table 3: Concepts or features that workshop ideas featured recurrently and that are explored in the DFs

| Concept/Feature | Role of AI | Design Fictions |
|---|---|---|
| Personalization | AI is used to modify the content or news selection to meet the needs and desires of the specific user. This may be done automatically or at the user's request. | DF1, DF2, DF4, DF5 |
| Content Transformation | AI is used to transform the mode, language or format of the content (e.g., from text to cartoon; from complex to basic language) | DF1, DF2, DF6 |
| Simulating a Person | AI is used to generate and control a virtual representation of a person (e.g., a virtual news presenter, or commentator) | DF1, DF5, DF6 |
| Gathering Information from Multiple Sources | AI gathers news from multiple sources for the user, potentially even from non-traditional sources | DF1, DF3 |
| Adjusting the Amount of Detail | AI is used to abstract, expand, or cut news items to match the user's interests or temporary needs | DF1, DF3, DF4 |

*4.1.1 DF1: Zenith*

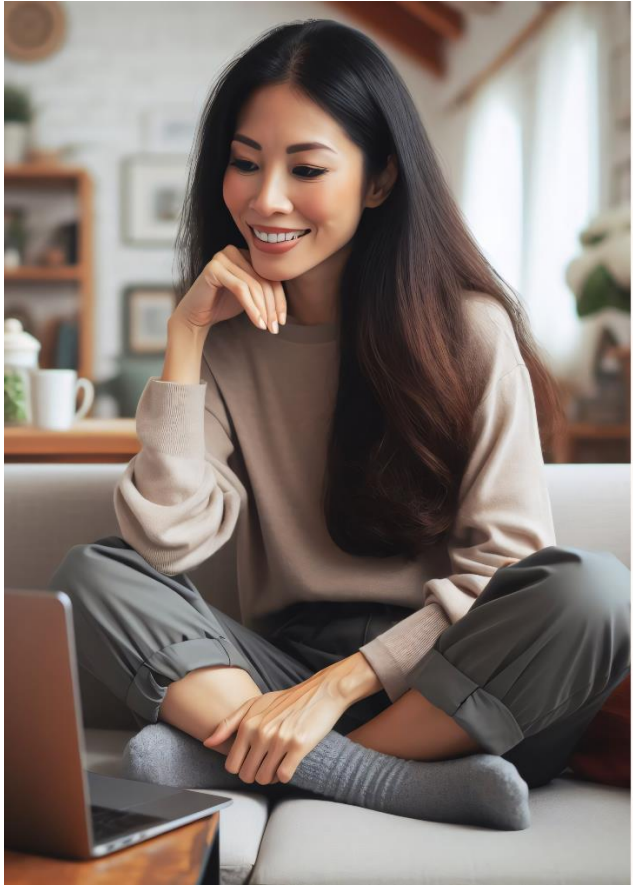

Figure 2: *Zenith* is an AI assistant that shares the news like a close friend.

The initial idea for *Zenith* (Figure 2) came from a workshop participant's story about their neighbor avoiding news completely due to feeling overwhelmed and anxious, while attempting to keep up to date by having conversations concerning recent events with other people. The DF depicts an AI assistant that delivers information in a friendly and upbeat manner, covering both personal and global news while being mindful of the user's emotional well-being. News avoidance, driven by perceptions of negativity, overwhelm, and untrustworthiness, presents a significant societal challenge [10]. However, *Zenith* also incorporates concepts that workshop ideas recurrently featured: personalization, collecting news from a wide range of sources, AI simulating a person, transforming news from one format to another (here to discussion format), and adjusting the amount of detail.

After creating *Zenith*, we noticed it shares similarities with Søndergaard & Hansens' [89] AYA and "U," and Rezwana & Maher's Design Pal [75]. AYA and "U" are feminist DFs of personal digital assistants (PDAs) [89]. AYA is a PDA that pushes against sexual harassment, for example, by contacting the harassing user's mother about how the user has called it. "U" is a PDA that gives birth control advice but makes a mistake, and the user accidentally becomes pregnant. Design Pal is a DF of a speaking, holographic, AI assistant that aids designers in product design [75]. Rezwana & Maher presented Design Pal in the form of a story about a student of design who uses it because they are running out of time with their assignment. Further, in science fiction, similar concepts to *Zenith* can be found in stories like "Her" by Spike Jonze, a sci-fi about a lonely man who develops a deep emotional relationship with an AI.

*4.1.2 DF2: NewsComics*

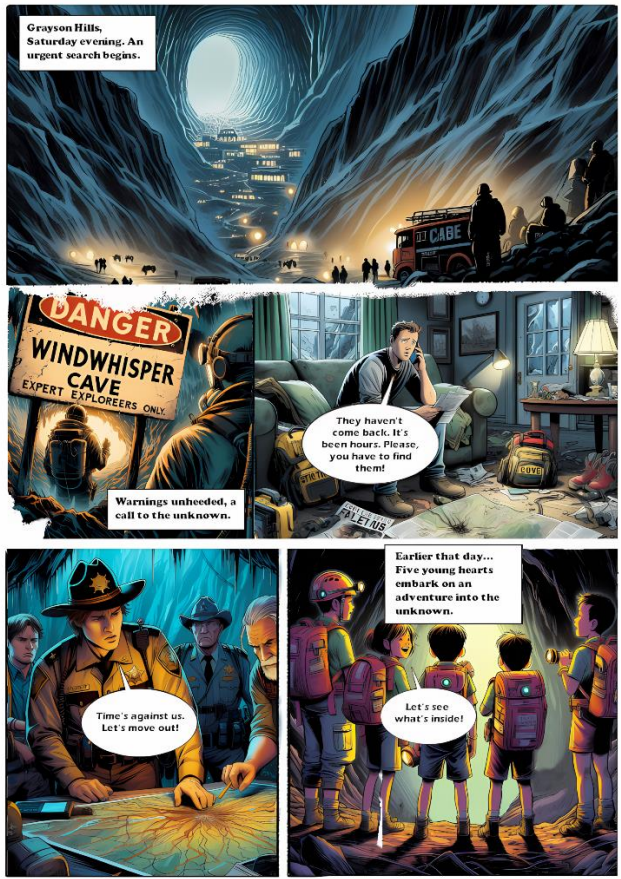

Figure 3: *NewsComics* creates personalized, AI-generated comic news for children to make news more engaging and accessible.

The recurring workshop ideas of AI-based, personalized news curation, and content transformation inspired the creation and selection of *NewsComics* (Figure 3). *NewsComics* is a service that transforms news articles into personalized, AI-generated comic books designed to make current events engaging and accessible particularly for children. The DF was also selected to highlight generative AI's ability to mix fact with fiction. Works like Feed by M.T. Anderson, where news is seamlessly integrated into a personalized, AI-driven feed, echo the potential future suggested by *NewsComics*. The use of comics to present news also touches on the notion of storification and the simplification of complex issues, a common theme in dystopian science fiction, where critical thinking is often sacrificed for entertainment. Further, the idea to present the fiction in the form of an advertisement came from previous work [7].

*4.1.3 DF3: Discover*

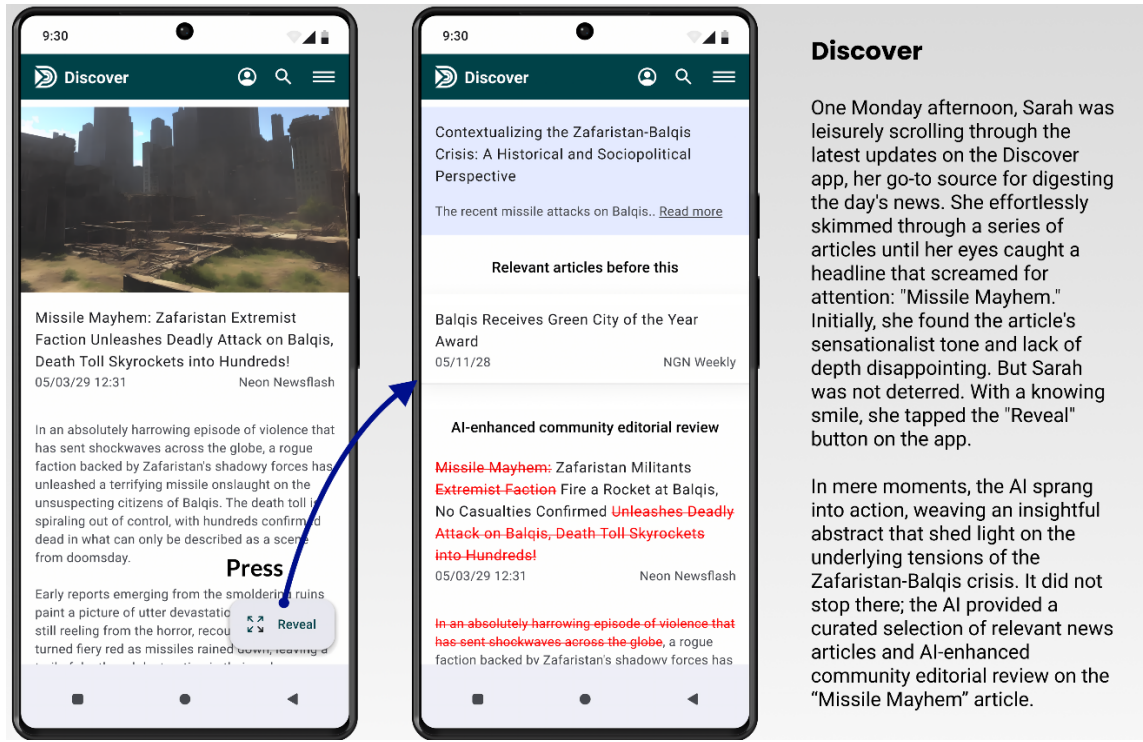

Figure 4: The *Discover* app uses AI to contextualize and review news. The illustration features the following resources: BRIX Templates, the smartphone frame; Artem K, the logo; Material Design, the UI kit.

Workshop participants had many ideas that featured AI adjusting the amount of detail in news content or AI gathering information from multiple sources. This, along with one participant's idea of using AI to put the news in context, inspired the creation and selection of *Discover* (Figure 4). The fiction illustrates an AI-driven mobile app that can contextualize news in three ways: by providing summaries that put events into context, presenting relevant news articles that precede the current news article, and providing AI-enhanced community editorial reviews. The idea that AI provides background information is similar to what human journalists already do [59]. Also, present-day tools like *TLDR This* offer AI-generated summaries of content. The idea of AI-enhanced community editorial review, however, is more novel. It is intended to highlight AI's ability to gather information from multiple sources, including from online communities. Additionally, Discover is futuristic in the sense that humans are likely still better at fact-checking than current AI systems, as recent studies have argued [9].

*4.1.4 DF4: NewsLens*

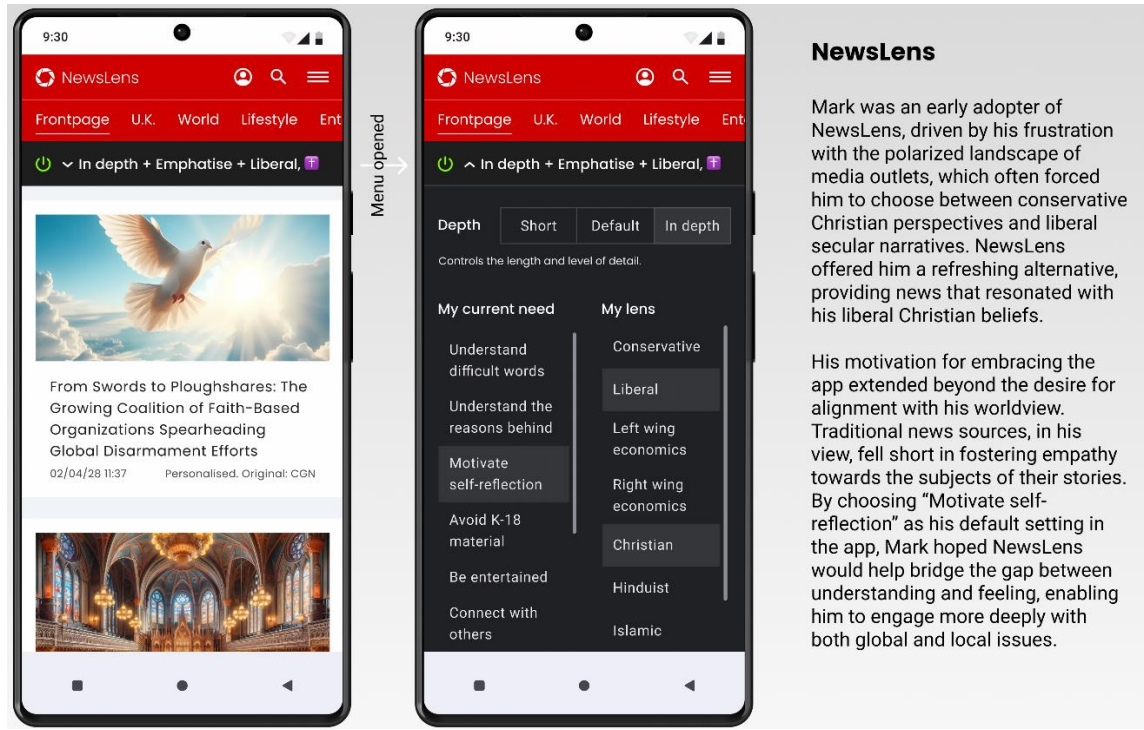

Figure 5: The *NewsLens* app uses AI to filter and modify news articles to match users' current needs and religious and political views. The illustration features the following resources: BRIX Templates, the smartphone frame; Artem K, the logo; Material Design, the UI kit.

*NewsLens* (Figure 5) was created and selected as the workshop ideas recurrently featured the concept of personalization, and the ideas often revolved around the wish to have more control over the news content consumed. *NewsLens* is a news app that allows users to filter and modify news articles to match their personal beliefs and values, as well as specific needs that they may have. While the presented UI is a mobile app, the scenario is realistic also on other device form factors. The DF was selected also because it might raise questions about what the user could realistically control about the news with the help of AI and how they could stay in control.

*4.1.5 DF5: Together*

*Together* (Figure 1) was inspired by recurring workshop ideas of personalized news presenters, personalization, and the first author's idea of using AI to make news consumption more social. *Together*, presented at the beginning of this paper, is a DF of an AI-enhanced smart TV that detects users' presence, dynamically adjusting the content and the virtual news anchor to cater to their individual or combined tastes. Material Design UI kit was used in the illustration. The DF may bring to mind the science-fiction film "Minority Report" directed by Steven Spielberg, where in one scene, the protagonist walks through a mall, and the billboards he passes identify him and display personalized advertisements. This level of personalization in media is similar to the AI-enhanced smart TV in *Together*.

*4.1.6 DF6: Forms*

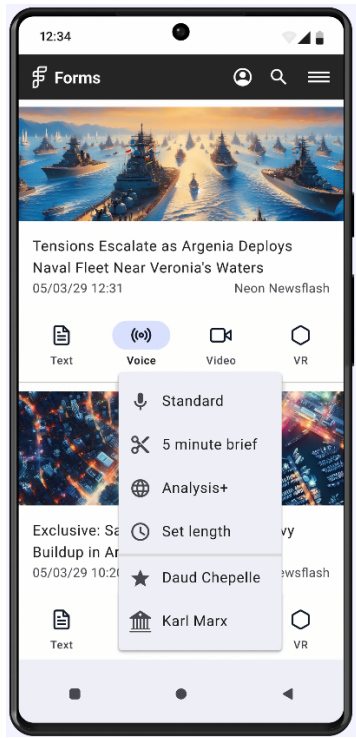

Figure 6: *Forms* is an AI app for converting news into podcasts and other formats. The user can choose the length and depth of the podcast. The app also simulates famous personalities as news commentators. The illustration features the following resources: BRIX Templates, the smartphone frame; Artem K, the logo; Material Design, the UI kit.

Recurring workshop ideas of personalization and simulating a human journalist are manifested in *Forms* (Figure 6). *Forms* is an AI app that transforms various types of news content into customized formats, like podcasts and VR. The app also offers options for controlling length, analysis depth, and even simulating famous personalities as news presenters and commentators. Additionally, Cameron's dissatisfaction with the app's pricing model and overabundance of features in *Forms* is reminiscent of themes found in The Hitchhiker's Guide to the Galaxy by Douglas Adams. Adams' satire often highlights the absurdities and inefficiencies in futuristic technologies and systems.

4.2   Analysis: Challenging the Assumptions Underlying the Fictions

This section details the results of phase 2 of the research, encompassing DF analysis workshops with experts and co-writing analyses with them. While in the previous section, we presented DFs of AI news applications that may become technologically possible, at this stage we critique the ideas with experts. Although many of the critical insights we present have been noted in the previous research we refer to, our results provide a nuanced perspective on how these issues materialize in concrete and plausible contexts of media consumption involving these technologies. We begin by challenging the assumptions that the design team made when crafting the imagined scenarios.

*4.2.1 Can an AI system reason like a journalist?*

Experts in the workshops expressed doubt about the capability of AI to achieve the levels of factual accuracy and newsworthiness people require for journalism. Specifically, they questioned how AI could manage not only the factual content of a news story but also the nuanced decisions about what to report and how to present it that consumers expect journalists to make. This concern is related to the idea of AI companions taking strong agency in news delivery and personal filtering (e.g., DF1 *Zenith*). It is noteworthy that journalists are guided by complex factors such as context, public importance, and ethical considerations—areas where AI models like ChatGPT remain limited [36, 37].

For example, one expert shared an analogy to underscore this challenge:

> "People need to know that they should close their windows if there is a power plant accident. This is presently [the national broadcaster's] responsibility. *Zenith* should know what needs to be communicated. ... *Zenith* would probably need to make quite a few logical connections to understand that if we have a major accident, it is important to communicate that clearly." (Expert #2)

While current AI can modify news content's tone and style, the lack of deep contextual and societal understanding means that it struggles to provide nuanced, contextually relevant information. This shortfall is especially critical when AI must communicate essential safety updates, as even small inaccuracies could have severe consequences.

Similarly, the experts questioned how the AI in DF4 *NewsLens* could adjust news items to fit a variety of perspectives and their combinations as requested by the user. They noted that while labels associated with political approaches and religious orientations can be distinguishable at an exemplary level, the reality is that these labels encompass an endless array of nuances and even internal contradictions. This could result in AI-generated news where perspectives are tweaked towards a particular direction, appearing shallow or even caricatured, as the experts observed:

> "Perhaps a left-right division might work in an American context, but what about all the other classifications? If it goes even slightly deeper than that kind of dichotomy, then the question of training material becomes critical. ... Also, [what if] you have a strong filter or lens selected but that filter does not work as expected, it can lead to a pretty strong negative experience." (Expert #5)

> "Imagine that the AI makes every picture in an article to feature someone with a headscarf, because somehow it has stereotyped Islam like that. It takes a person who has grown up in that world to know when it is their own perspective and not a forced stereotype." (Expert #1)

Besides worrying that AI would oversimplify issues and misrepresent people when tweaking the perspectives, the experts further underlined that reconstructing a news article from a different foundational perspective is an even more difficult challenge. This process would involve more than just adjusting language or emphasis; it requires a deep understanding of the underlying beliefs, values, and cultural contexts that shape the perspective. Any failure in this nuanced understanding could lead to a distortion of the original message, potentially alienating readers and eroding trust in the AI's capabilities.

*4.2.2 Can people accept integrating AI into their daily lives?*

In the workshops, the experts raised significant concerns about whether AI applications described in the DFs could realistically gain widespread acceptance among consumers and audiences. AI-driven personalization of news content, both in recommendations and the content itself, along with the simulation of human behavior (as seen in DF1, DF5, and DF6), would require people to allow AI to become more integrated into their daily lives in a number of ways. These applications would necessitate not only an increase in the production of detailed data about consumers' private lives but also AI assuming roles that are perceived as deeply personal—such as news announcers or personal AI assistants.

If introduced today, such AI technologies would face a climate already marked by scepticism toward both technological advancements and traditional journalism. Recent studies indicate that over half of the global population lacks trust in AI, with privacy violations, data misuse, and a lack of transparency in how personal information is handled cited as major concerns [32]. Surveys also show that most people are more comfortable with news created by human journalists than with AI-generated content [23]. Furthermore, trust in established institutions, including government and media, has declined, particularly in developed nations [72].

Experts observed that in the fictional worlds of DF5 and DF1, public perceptions of data surveillance seem to have shifted in comparison to the present. Unlike the current environment, where surveillance is often an unseen background process, the AI applications in these fictions make it an explicit and integral part of daily life, using the data they collect in real time to shape interactions. The experts suggested that for users to engage effectively with such technologies, their attitudes toward data surveillance would need to be highly permissive, perhaps even more accepting of its pervasive and intrusive nature than at present. One expert highlighted this shift:

> "[Considering what could have made DF5 *Together* possible] perhaps there has been a change in terms of [people] accepting surveillance or the idea of how different data-collecting systems are present. If we think about the fact that the TV recognizes the viewers ...the fact they would know what news you want to consume at each moment alone is concerning. [Think about] the amount of information about you that would be generated—who you live with or who you no longer live with, and for instance, what your political opinions are." (Expert #3)

The experts noted that a separate but similarly personal development seems to have happened in relation to the virtual human characters present in DF1, DF5, and DF6. They pointed out how the users in these DFs appear to have moved beyond the tendency to prefer human interaction over artificial counterparts documented in recent research [25, 51], or perhaps are comfortable with interaction with AI because of a lack of human interaction [8]. In DF5, for example, users are encouraged to abandon the notion that a newsreader must be a human being. This reflects research on parasocial interaction with newsreaders, where viewers have been seen forming emotional connections with media figures, even when they are not human [80].

*4.2.3 Can news be storified for engagement without misleading the audience?*

In discussing DF2 *NewsComics*, experts raised concerns about the feasibility of creating engaging comics or other forms of engaging presentations from traditional news stories, particularly when covering complex or policy-heavy topics. While policy proposals can have a substantial impact on daily life, they do not easily lend themselves to comic-style storytelling. For example, one expert observed:

> "If the news topic is extremely boring, it would need to be turned into an engaging story ..but at the same time it would need to be clear to readers—from the perspective of [the local journalists' ethical guidelines]—when something has been dramatized.." (Expert #1)

This led to questions about whether the shift to news comics might inadvertently cause a change in focus from hard news, which is fact-based and timely, to soft news, which tends to be more entertaining and centered on human interest or lifestyle. Prior research on AI-driven content, such as voice assistants, has already shown a trend toward lower information quality when dealing with hard news compared to soft news [15]. Similar challenges could arise in generating news comics from more complex news stories.

*NewsComics* is primarily targeted at children, which sparked mixed reactions among the experts. They noted different efforts of media companies to produce content especially aimed at children, such as The New York Times for Kids. However, transforming general news into a comic format that resonates with children requires both creativity and a nuanced understanding of what will captivate their attention. After extended discussion, the experts agreed that while humor and creativity can indeed make a mundane news story more engaging, these elements often come at the expense of accuracy. One expert proposed a compromise—rather than adapting specific news events into comics, the focus could shift to producing educational comics that address broader news themes:

> "[*NewsComics*] could be toned down to focus on addressing a current theme, which would give it an educational purpose, and perhaps even teachers could use it in schools. ... I like the idea of abstracting general topics for children. It is educational; it would teach them to recognize general phenomena and trends." (Expert #5)

The experts appeared to think this approach could balance the need to engage children while maintaining educational value, helping young readers understand broader societal issues without oversimplifying or distorting the facts.

*4.2.4 Are news consumers as rational truth-seekers as imagined?*

The experts pointed out that the DFs seem to operate on the assumption that news consumers are primarily rational truth-seekers. This assumption is most evident in DF3 *Discover*, where users are expected to be motivated by a desire to explore the broader context of a news article and seek out deeper truths. However, the experts expressed doubts about this premise, suggesting it may be overly idealistic. For example, one expert remarked:

> "Is it wrongly assumed that the user is a rational truth seeker? Why should *Discover* operate on a rational basis—why not as something like a tabloid, for instance, featuring juicy gossip about politicians?" (Expert #7)

Similarly, the experts criticized DF4 *NewsLens* for relying on what they perceived as an elitist perspective. This design assumes that users are willing to consider diverse viewpoints thoughtfully while browsing news, implying a high level of engagement and intellectual effort. In reality, many users may not be inclined to engage in such reflective practices. Instead, they may often favor quick reads that confirm their existing beliefs, rather than investing the time required for a more nuanced, in-depth analysis of multiple perspectives.

Building on this, the experts discussed the significant challenges in forming a community dedicated to reviewing news content, as envisioned in DF3 *Discover*. The main issues they mentioned include attracting a large, active user base and preventing vandalism, where users might intentionally distort or sabotage content. While Wikipedia manages this issue with a combination of moderation and automated systems [13], scaling such safeguards to ensure the integrity of news content would be far more complex. The experts also warned of the possibility that users might manipulate the system to introduce biased or misleading information, which could erode the platform's credibility and compromise its goal of delivering accurate and trustworthy news assessments.

### 4.3 Analysis: Recognized Potential and Risks

In this section, we briefly highlight the potential positive societal impacts of the imagined systems before focusing on the more extensive societal risks and downsides associated with their use.

*4.3.1 Perceived benefits*

The experts acknowledged several potential benefits of the depicted applications. Many of them were seen as potentially reaching audiences that presently avoid or consume less news, for example, by attracting the attention of young consumers with novel ways of presenting timely content (DF2). In particular, the applications involving automatic personalization of news recommendations (DF1, DF5), were seen as helping to address the growing issue of information overload [40] and could potentially reduce news avoidance by helping users navigate vast amounts of information more easily. For example, the experts noted:

> "As long as Zenith ensures that the essential core, which enables a proper understanding of the matter, is covered, this could allow each user to receive just the right amount of information—and in the order they can absorb and find sufficient." (Expert #1)

> "[Together is motivated by] information overload and the challenge of so much happening in the world that it's hard to keep track of everything." (Expert #3)

Second, applications that provide in-depth analysis (DF3, DF4, DF6) were praised for their potential to foster a deeper understanding of complex issues, thus contributing to a more informed public. The experts also recognized the value of features that offer additional context to news stories (DF3) or allow users to explore alternative political perspectives (DF4), especially in an era of biased media coverage and political agendas. These tools could help create a more diverse and balanced media landscape. For example, it was observed:

> "Contextualization would be even more important in complex societal processes (e.g., political decision-making, regulatory operations, etc.), although it does help to provide context for the conflict here [in the Discover app]. ... Contextualization is a good idea, and it works best when it offers factual information (so-called fact boxes) as background." (Expert #2)

> "At its best, the [NewsLens] app could challenge the kind of polarized news landscape described. It could support news consumers in better understanding underlying reasons or in developing critical or empathetic thinking." (Expert #3)

As the experts recognized the value in these applications, they also noted specific improvements that would augment such value. For *Discover* (DF3), they recommended that the AI explicitly identify whose perspective on the conflict is being presented and what the sources are. Considering the video, podcast, and VR features of *Forms* (DF6), they suggested including clear disclaimers when fictional elements are used for illustrative purposes. They also suggested incorporating topic-specific lens options to *NewsLens* (DF4), including an option for lens randomization. For instance, news about an accident involving a car and children on an electric scooter could be analyzed through various perspectives, such as urban planning, traffic laws, or socioeconomic factors like poverty.

However, despite the potential positive effects and the specific improvements suggested, the experts emphasized the risks associated with these systems that appeared to far outweigh the benefits.

*4.3.2 Disconnecting people from shared reality*

One of the major concerns raised in the workshops was the impact of AI-driven personalization on the type and quality of information users receive. This concern touches on two key issues: the tension between engaging content and traditional journalistic ideals and the risks of selective exposure to information.

The first issue centers on how AI's ability to make content more engaging could blur the line between objective reporting and subjective interpretation. The experts worried that, in strongly tailoring content to user preferences, there is a risk that facts might be overshadowed by material designed to entertain or satisfy user preferences. This shift could undermine journalistic integrity, a risk that was seen to be present already with current news feed curators and clickbait journalism. Even when factual accuracy is preserved, the experts were concerned that entertainment and engagement might take precedence over content relevance. As one expert noted:

> "[What the user does in DF3 *Discover*] raises an important question: What exactly are we looking for when seeking reliable information? Often, we do not just want plain, unembellished facts; instead, we are drawn to content that is both engaging and relevant, and this makes these apps scary." (Expert #4)

DF5 *Together* and DF1 *Zenith* were specifically highlighted as examples of how AI personalization could create a future where entertainment and user satisfaction overshadow objective and comprehensive news coverage. These DFs assume users are rational truth-seekers, but the experts argued that this assumption might lead to the development of systems that, while claiming to provide in-depth information, instead offer highly personalized and potentially narrow content pathways. One expert's observation about DF5 *Together* underscored this concern:

> "People's understanding of society and its issues would narrow if, for example, those not interested in politics excluded it from their news or chose to follow only news that covers conflicts from a specific perspective or supports a particular political viewpoint. Or if someone's news consisted only of sports news." (Expert #3)

This comment ties into the second key issue: the potential for AI-driven personalization to exacerbate selective exposure to information. This phenomenon has long been associated with algorithmic targeting and content filtering, commonly referred to as "filter bubbles" and "echo chambers" [22, 71]. If users are primarily exposed to information that aligns with their pre-existing beliefs and preferences, their understanding of broader societal issues could become increasingly distorted. The experts emphasized that this selective exposure, already a concern in journalism (see [82]), could further undermine a shared reality by creating divergent content streams for different individuals.

*4.3.3 Nurturing disengagement*

The experts noted that in the fictional worlds presented, users may lose the ability to engage with challenging content, which could have negative long-term effects. Specifically, in the context of DF1 *Zenith*, they raised the concern that consistently filtering out negative news or difficult emotions could prevent individuals from building the emotional resilience needed to cope with real-world adversity. Emotional resilience is often developed through encountering and processing negative experiences [69], and a constant avoidance of such content might undermine this growth. The experts noted concerns:

> "Citizens will not learn to deal with negative things and emotions if they are never exposed to them." (Expert #2)

"If I allowed someone to constantly emphasize only the pleasant things, would I gradually lose some of my ability to take in information that does not necessarily interest or please me?" (Expert #8)

A similar concern arose during discussions of DF2 *NewsComics*. Several experts worried that turning news into more entertaining formats might hinder audiences—especially younger ones—from developing the capacity to engage with traditional, text-heavy journalism. This shift could lead to an over-reliance on simplified, bite-sized content. As one expert cautioned:

"This increases the phenomenon of everything turning into entertainment, which grows a generation of news readers who cannot handle long news articles or perhaps text-based news at all." (Expert #2)

Expert #5 further argued that the notion of *NewsComics* serving as a gateway to journalistic content for children is both idealistic and based on a consequentialist moral standpoint. They noted that such ideas appear common in app development, where new innovations are justified by their expected long-term benefits; but in reality, these developmental pathways are often based on a positivism bias towards technology and an oversized faith in the human capability to anticipate future developments.

*4.3.4 Decreasing diversity*

The experts also expressed concerns that some AI-driven applications, particularly those involving personalized news (DF1, DF4, DF5) and customizable news presenters (DF5), could inadvertently reduce diversity in media.

For example, in DF5 Together, while allowing users to customize the appearance of their news presenter, including characteristics like skin color and personality, could be thought to improve representation for minority groups, users' ability to do this could reinforce existing social divisions rather than promote inclusivity. If individuals consistently choose presenters who resemble themselves demographically, they may be exposed to fewer perspectives from outside their own social, ethnic, or political groups. As one expert explained:

"Choosing a news anchor could provide opportunities for minorities to have their news read by an AI representing their minority, but at the same time, diversity might actually decrease if the majority population were to make similar choices." (Expert #3)

This concern was echoed in discussions of DF1 *Zenith* and DF4 *NewsLens*. By allowing users to customize the types of news they receive based on their preferences, these apps could unintentionally narrow the scope of content, filtering news through specific political, religious, or socioeconomic lenses. Such personalization risks reinforcing users' existing biases and deepening societal divides. For instance, an AI system that tailors content to users with extreme or prejudiced views could exacerbate these biases, as seen in research on algorithmic "rabbit holes" [65].

*4.3.5 Reinforcing flaws of the media ecosystem*

The experts also highlighted a troubling aspect of DF3 *Discover*. The DF seems to assume the coexistence of both high-quality and low-quality content. They noted that the DF seems to depict a future where an advanced AI system can identify and correct misinformation. However, despite this technological advancement, unreliable news continues to spread widely. For example, one of the experts noted:

"Someone takes a [revised] article from here, runs it through some language model with 'make this into a sensational clickbait trash,' then someone feeds that article back here, and once again, the same system fixes the article. It's like they've invented a perpetual motion machine." (Expert #4).

On the other hand, the experts expressed some suspicion over whether the scenario would eventually materialize: the fiction paradoxically suggests that the user happily participates in this cycle by intentionally choosing to read sensational and factually dubious content and *then uses AI* to rectify errors and reduce sensationalism.

In the case of DF6 *Forms*, the experts also expressed concerns about how integrating AI with paywalls could widen the information gap between different socioeconomic groups. AI could make premium content more appealing and accessible to paying subscribers, while non-subscribers would be left with lower-quality news, potentially laden with intrusive ads. This could further entrench a two-tiered media system, where those without access to premium content do not receive critical information that shapes public discourse and decision-making.

## 5 DISCUSSION

The results of this study highlight potential benefits but also significant ethical and societal concerns surrounding the integration of AI into news consumption. The following discusses the implications of various findings for the design of AI-infused apps for news consumption and reflects on the methodology.

### 5.1 The need for journalistic insight in AI development

The DF analysis workshops expressed skepticism about AI systems' ability to deliver news in a thoughtful and nuanced manner. AI, while capable of mimicking journalistic output and could successfully deliver facts, still falls short in areas requiring deep contextual understanding, ethical decision-making, and cultural sensitivity customary to journalistic processes [28, 36, 37]. These doubts can be motivated by an understanding of the demanding journalistic work and production of news content. Journalists often describe their practices as guided by strong professional ethics and emphasize human responsibility concerning algorithmic and AI technologies introduced in the production and distribution of content [81]. Similarly, journalism researchers suggest that news coverage is to meet high expectations, such as addressing public needs, values, and sensitivities [99].

Furthermore, the workshop discussions highlighted the tension between creating engaging news content and maintaining journalistic ideals of objectivity and truth. The DFs suggested that AI-driven news might prioritize user engagement over factual accuracy, raising ethical concerns about its influence on public discourse. In particular, the risks of oversimplification and misrepresentation in AI-generated news, such as in DF4 *NewsLens*, were emphasized. This underscores a key challenge in HCI: designing interfaces that present information accurately while managing cognitive load [67]. Additionally, AI's role in modifying news may create conflicts of interest, with owners prioritizing engagement over accuracy, which is something that, for example, Facebook's owners have been accused of [97].

This underscores the imperative for developers and journalists to work closely together to responsibly design and deploy AI-enhanced news applications. This reflects classical Human-Centered Design principles [11] of involving end-users and stakeholders—in this case, journalists—throughout the process. Earlier research has suggested that developers need journalists' deep understanding of journalistic values to design AI systems that perform journalistic tasks [45]. This outcome echoes previous findings in data journalism [74, 93], from where further best practices could be examined. The similarity of these findings echoes the special function journalism has in society and thus the demand to support its operation model. Following this, collaboration with both journalistic professionals and researchers could help identify existing and emerging risks in the deployment of AI technologies in media production and consumption. Moreover, this work can be informed by insights from the development of recommender systems [21, 76] as well as extant results on the ethical and societal implications of the AI and algorithmic technologies in the distribution of media content [83]. As argued in recent research, the significance of these developments centrally depends on our conceptualization of the central values and norms at stake, such as (human) autonomy (Rydenfelt et al. 2025). Speculative design, such as conducted in this research in terms of design fictions, could be used to prompt shared reflection by journalists, researchers and developers on the ethical and societal implications with respect to new AI applications at different stages of development.

### 5.2 Promoting diversity in AI-driven news

Based on the workshops, the potential for AI applications to decrease diversity poses a number of design challenges, as recently identified by a meta-review of previous research [86]. One constant concern with new technologies in the distribution of media and journalistic contents is the potentially selective exposure to content, including reduction in the diversity of topics and perspectives in individual consumption [82]. Recent studies suggest that user satisfaction with content recommendations depends on both accuracy and diversity [42, 87].

Integrating diversity into AI-driven news systems to meet the concerns voiced particularly over the DFs featuring highly personalized news and media content (DF1, DF4, DF5) requires strategies that may draw from the results of recent research for the curation of content that encompass varied cultural, socioeconomic, and political contexts and perspectives. For example, research highlights that bias in training data contributes significantly to underrepresentation of minority viewpoints [3, 63]. Similarly, incorporating socioeconomic diversity in datasets helps to ensure that reporting reflects different population experiences, reducing stereotypes and stigma against marginalized groups. However, it is also important to recognize that the GenAIs currently available have demonstrated biases [53, 104]. For instance, a recent study analyzing images produced by popular GenAI tools—Midjourney, Stable Diffusion, and DALLE 2—found systematic gender and racial biases [104]. Further, recent computational journalism research highlights that while advances have been made in developing algorithms that can detect and address framing biases, identifying content patterns that favor certain perspectives, the task is far from being solved [78].

Designers should consider how customization features, like those in DF5 *Together* may reinforce existing biases, and how such features may require balancing with strategies to promote diverse representations. These could include promoting inclusive defaults and implementing features such as suggestions to explore presenters with different backgrounds.

Additionally, access barriers, such as paywalls, referred to in DF6 *Forms*, limit underserved communities' access to high-quality information [95]. To address these concerns, innovative access models, such as subscription waivers for institutions in low-income regions or publicly funded repositories could provide broader access.

### 5.3 Reflection on the Methodology

One of the main challenges of this study was determining the scope of the DFs—specifically, what elements to include and how many fictions to present. We concluded that focusing on just one or two fictions would be less effective than our broader approach, as each DF sparked new ideas and discussions. Additionally, creating a single fiction with all the AI functionalities seemed impractical and too unrealistic. Another consideration was whether to adjust the fictions after analysis workshops, but despite some credibility issues, the fictions effectively stimulated reflection and engagement. Therefore, we chose to keep them as is, letting their fictionality provoke thought.

Certain limitations in our methodology should be acknowledged. First, the relatively small number of workshop participants and the relatively low number of design fictions limit the diversity and extent of perspectives present in the data. Nevertheless, the results indicate that several critical insights emerge from the analysis. The speculative DFs proved valuable as boundary objects, allowing the surfacing of potential issues with AI-enhanced news applications and stimulating meaningful conversations both within the design team and in the workshops. At the same time, it is to be emphasized that the open-ended nature of speculative design necessitates cautious interpretation.

Second, regarding our positionality, the experts' views and values align with liberal-democratic ideals and journalistic principles, such as objectivity, truth, and an interest in education. Their professional backgrounds in journalism and technology significantly influenced both the creative and analytical dimensions of our work. Insights from Journalism and Media Studies directed our focus on AI's impact on news consumption, while expertise in Computer Science and Human-Computer Interaction brought technical, design-oriented, and user-centered perspectives to the table. For instance, expertise in Journalism and Media Studies played a critical role in ideation, selecting themes, and calibrating the speculative nature of the design fictions. Additionally, ethical insights from experts familiar with journalistic ethics (e.g., [98]) and IT ethics enriched our consideration of user agency, privacy, and responsible reporting during the analysis workshops.

Importantly, the experts did not remain confined to their professional expertise; they assumed diverse roles as subject matter experts, designers, content producers, news consumers, and technology users. This multifaceted engagement deepened their contributions, reinforcing findings from similar studies (e.g., [101]). Moreover, the familiarity among some participants fostered richer discussions, while the fresh perspectives of newer members stimulated debate and challenged assumptions.

Finally, while our over three-hour-long DF analysis workshops enabled in-depth engagement and comprehensive reflection, they also resulted in participant fatigue. This setup contrasts with previous studies, such as Wong et al. [101] and Rezwana & Maher [75], where experts engaged with DFs for shorter, more concentrated periods. Despite the demanding nature of our extended sessions, the experts' willingness to participate underscores the significance of the subject matter and the effectiveness of the fictions in stimulating critical reflection. In hindsight, our iterative and collaborative approach helped balance potential biases and address participant limitations, enhancing the overall robustness of our findings.

## 6 CONCLUSION

To get ahead of the curve in understanding the opportunities and risks posed by AI in news and media consumption, this study presented six imaginative, yet plausible design fictions featuring cutting-edge generative AI features to enrich news consumption. The fictions showcased intriguing, yet possibly troubling features from hyper-personalized news recommendations to format-shifting content and AI-powered virtual news anchors. Through workshops and collaborative writing with experts in journalism, media studies, philosophy and the study of technology, these scenarios were critically examined, revealing many reasons for caution. The analysis underscores the necessity for AI systems to navigate the delicate balance between user engagement and the foundational principles of objectivity and truth in journalism. Although many of the critical insights and opportunities revealed in our results may be somewhat familiar to readers, including designers of AI systems and applications, our findings offer a richly detailed perspective on how these issues emerge in concrete and multifaceted ways in plausible contexts of introducing these technologies into media consumption. In this way, our results also outline the potential of employing similar methodologies to chart further challenges and opportunities. The future of AI in news may be bright—but only if we tread carefully.

ACKNOWLEDGMENTS

The work was funded by the Helsingin Sanomat Foundation and the Kone Foundation. We thank all the ideation workshop participants.